L. Mello# Fat-Tailed Distributions in Catastrophe Prediction

Louis Mello

The notion that natural disasters can be controlled is, of course, farcical; history is permeated with examples of countless failed attempts at this pointless task; it is synonymous with trying to build a perpetual motion machine. Nonetheless, there are ways to reduce their impact on human communities, particularly by looking away from the *normal* hypothesis.

In a press conference on the remnants of Katrina, the commander of the Army Corps of Engineers, Gen. Carl Strock, asserted: "… when the project was designed -- … we figured we had a 200 or 300 year level of protection. That means that the event we were protecting from might be exceeded every 200 or 300 years. That is a 0.05% likelihood. So we had an assurance that 99.5% of this would be okay. We, unfortunately, have had that 0.5% activity here."

This argument operates under two assumptions:

1. Given that this issue is entirely based on probabilities, there is no assurance of anything at all.
2. The estimate presented is based on a Gaussian bell shaped curve, the proverbial *Normal Curve.*

Since the late 1800s, researchers have been aware that the probability of what are called "extreme events", i.e., events that fall on the tail ends of a statistical distribution and, as such, are the most likely **not** to occur, cannot always be accurately described by the bell-shaped curve. Such manner of activity is usually much more appropriately described using "fat-tailed" or the stable-Paretian class of distributions.

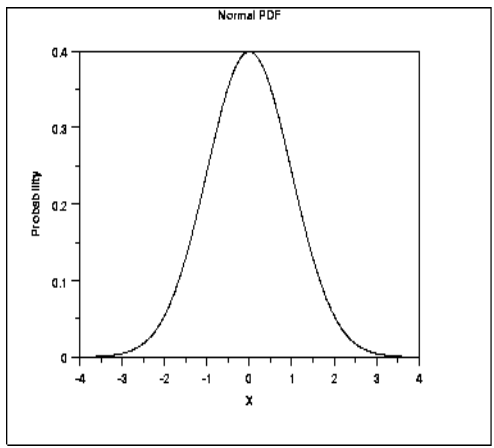

*An example of a curve with "fat tails" is the Cauchy distribution (a member of the stable-Paretian class), shown below. In the case of the normal curve (on the left), the tails approach zero at -3.5 and 3.5 standard deviations. In the case of the Cauchy distribution the curve is still not close to zero at -5 and 5 standard deviations. This illustrates the higher probability at the tail ends.*

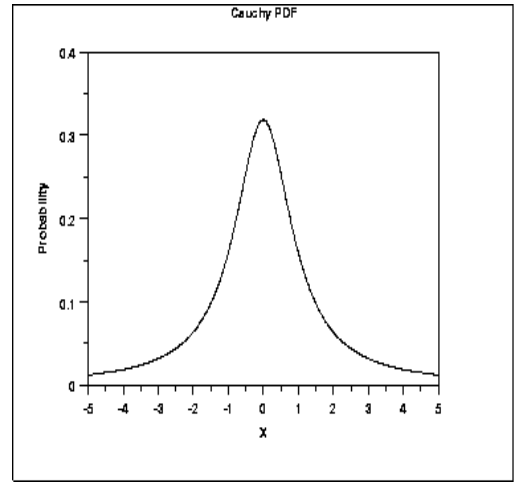

In its original form, this class of curve hails back to the Italian economist Vilfredo Pareto, one of the leaders of the Lausanne School. He originally proposed it to describe allocation of wealth, given how well it illustrated the mode by which a larger portion of the means in any society is owned by a very small percentage of the population. This same distribution has now recently found applications in fields as varied as finance (i.e., the return and volatility of risky assets), flood control (the variation of water levels in natural formations) as well as in the propagation of wildfires in densely wooded terrain.

In essence, under a Pareto type distribution, the probability of an "extreme event" is higher than it would be under the *normal* hypothesis. Hence, the 0.05% mentioned above could well have been closer to 5% and the "best case" scenario, 300 years, might have been reduced to only 63 years.

The Gaussian curve has long been the flagship of statistical analysis given the relative ease of computation and the overly abundant literature that exists on the topic. Fat-tailed distributions, on the other hand, have no mathematically closed form, making calculations much more taxing and more grudgingly reliant on simulatory technique. In recent years they have received an overstated degree of attention, by and large because of their possible applications in the universe of speculative markets; however, they remain resistant to many of the analytical methods within which the Gaussian can more easily be framed.

## *Mathematical Considerations on Fat-Tailed Distributions*

The mathematical depiction of the generalized characteristic function for the fat-tailed distribution is given by:

$$\log(f(t)) = i\sigma t - \gamma |t|^\alpha \left(1 + i\beta \frac{t}{|t|} \tan\left(\alpha \frac{\pi}{2}\right)\right) \qquad (0.1)$$

$\sigma$ = the location parameter of the mean.

$\gamma$ = is the scale parameter to adjust differences in time frequency of data.

$\beta$ = is the measure of skewness with $\beta$ ranging between -1 and +1.

$\alpha$ = the kurtosis and the fatness of the tails. Only when $\alpha$ = 2 does the distribution become equal to the Gaussian distribution.

The estimation of $\alpha$ is the most important factor in determining how the probability density function (pdf) will behave. The most accurate way currently known to estimate this parameter is by way of the Hurst exponent ($h$), whose relation to $\alpha$ is given by:

$$\alpha = \frac{1}{h} \qquad (0.2)$$

The *h* parameter can be estimated through the logarithmic regression of the Rescaled Range, defined by Mandelbrot[1] as:

$$R/S = aN^h \qquad (0.3)$$

where $a$ = a constant, $N$ = the total number of data points in the times series.

Obviously, when $h = \frac{1}{2}$ the time series would be a random walk, as such its associated probabilities would best be estimated through a judicious utilization of the normal distribution. The study of catastrophes (especially floods, as researched by E. Hurst[2]) has shown that *h* values usually range from 0.7 to as much as 0.9. This means, in effect, that the time series of this type of data possesses short-term memory; past data influences the immediate future and hence, the probabilities associated with this distribution follow a fat-tailed model as opposed to the Gaussian distribution.

For the special class of the original Pareto distribution we have a much more simplified distribution function:

$$1 - \left(\frac{\sigma}{x}\right)^{\alpha} \qquad (0.4)$$

It is easy to glean the close relationship between (0.4) and (0.1). And again, the shape or kurtosis parameter is the most important in the determination of the associated probabilities. These facts clearly show that only under a more robust form of distribution function can the effectiveness of catastrophe prevention be enhanced, i.e. take under consideration a more ***real*** probability estimate of extreme events.

---

[1] Mandelbrot, Benoit, "*The Fractal Geometry of Nature*", W.H. Freeman, 1981
[2] Hurst, H. E. "*The Long Term Storage of Reservoirs*", *Transactions of the American Society of Engineers,* 116, 1951